\documentclass[fleqn,usenatbib]{mnras}

\usepackage{newtxtext,newtxmath}
\usepackage[T1]{fontenc}
\usepackage{ae,aecompl}

\usepackage{graphicx}	% Including figure files
\usepackage{amsmath}	% Advanced maths commands
\usepackage{amssymb}	% Extra maths symbols
\usepackage{subfig}
\usepackage{CJK} %Chinese
\usepackage{pdfpages}

\usepackage[flushleft]{threeparttable}

\title[Near-infrared variability in dusty white dwarfs]{Near-infrared variability in dusty white dwarfs: tracing the accretion of planetary material}

\date{Accepted XXX. Received YYY; in original form ZZZ}

\pubyear{2019}

\begin{document}
\label{firstpage}
\pagerange{\pageref{firstpage}--\pageref{lastpage}}

\begin{CJK}{UTF8}{gbsn}
\author[L. K. Rogers et al.]{
Laura K. Rogers$^{1}$\thanks{E-mail: laura.rogers@ast.cam.ac.uk},
Siyi Xu (许\CJKfamily{bsmi}偲\CJKfamily{gbsn}艺)$^{2}$,
Amy Bonsor$^{1}$,
Simon Hodgkin$^{1}$,
\newauthor Kate Y. L. Su$^{3}$,
Ted von Hippel$^{1,4}$, 
Michael Jura$^{5}$\thanks{Deceased}
\\
% List of institutions
$^{1}$ Institute of Astronomy, University of Cambridge, Madingley Road, Cambridge CB3 0HA, UK \\
$^{2}$ Gemini Observatory, 670 N. A'ohoku Place, Hilo, HI 96720, USA \\
$^{3}$ Steward Observatory, University of Arizona, 933 N Cherry Avenue, Tucson, AZ 85721, USA \\
$^{4}$ Department of Physical Sciences, Embry-Riddle Aeronautical University, Daytona Beach, FL 32114, USA \\
$^{5}$ Department of Physics and Astronomy, University of California, Los Angeles, CA 90095-1562, USA
}

\maketitle

\begin{abstract}
%This is a simple template for authors to write new MNRAS papers. The abstract should briefly describe the aims, methods, and main results of the paper. It should be a single paragraph not more than 250 words (200 words for Letters). No references should appear in the abstract. 

The inwards scattering of planetesimals towards white dwarfs is expected to be a stochastic process with variability on human time-scales. The planetesimals tidally disrupt at the Roche radius, producing dusty debris detectable as excess infrared emission. When sufficiently close to the white dwarf, this debris sublimates and accretes on to the white dwarf and pollutes its atmosphere. Studying this infrared emission around polluted white dwarfs can reveal how this planetary material arrives in their atmospheres. We report a near-infrared monitoring campaign of 34 white dwarfs with infrared excesses with the aim to search for variability in the dust emission. Time series photometry of these white dwarfs from the United Kingdom Infrared Telescope (Wide Field Camera) in the \textit{J}, \textit{H} and \textit{K} bands were obtained over baselines of up to three years. We find no statistically significant variation in the dust emission in all three near-infrared bands. Specifically, we can rule out variability at $\sim$\,1.3\% for the 13 white dwarfs brighter than 16th mag in \textit{K} band, and at $\sim$\,10\% for the 32 white dwarfs brighter than 18th mag over time-scales of three years. Although to date two white dwarfs, SDSS J095904.69$-$020047.6 and WD 1226+110, have shown \textit{K} band variability, in our sample we see no evidence of new \textit{K} band variability at these levels. One interpretation is that the tidal disruption events which lead to large variabilities are rare, occur on short time-scales, and after a few years the white dwarfs return to being stable in the near-infrared.

\end{abstract}

\begin{keywords}
white dwarfs -- circumstellar matter -- infrared: planetary systems -- methods: observational -- techniques: photometric
\end{keywords}

%%%%%%%%%%%%%%%%%%%%%%%%%%%%%%%%%%%%%%%%%%%%%%%%%%

%%%%%%%%%%%%%%%%% BODY OF PAPER %%%%%%%%%%%%%%%%%%

\section{Introduction}

% This is a simple template for authors to write new MNRAS papers. See \texttt{mnras\_sample.tex} for a more complex example, and \texttt{mnras\_guide.tex} for a full user guide.

%All papers should start with an Introduction section, which sets the work in context, cites relevant earlier studies in the field by \citet{Others2013}, and describes the problem the authors aim to solve \citep[e.g.][]{Author2012}.

%%%%%%%%%%%%%%%%%%%%%%%%%%%%%%%%%%%%%%%%%%%%%%%%%%

The atmospheres of polluted white dwarfs have been interpreted as evidence for the survival of outer planetary systems to the white dwarf phase. Observations suggest that between 25--50\% of white dwarfs display elements heavier than helium in their atmospheres \citep{zuckerman2003metal, zuckerman2010ancient, koester2014frequency}. The rapid gravitational settling times in comparison to the white dwarfs' cooling age implies ongoing accretion from a reservoir \citep{koester2009accretion}.

The favoured theory regarding the pollution source is that it originates from planetesimals scattered on eccentric orbits towards the white dwarf where they tidally disrupt, and subsequently accrete on to the atmosphere of the white dwarf. These planetesimals have masses comparable to solar system asteroids \citep{debes2002there, jura2003tidally, farihi2010rocky, jura2014extrasolar, veras2014formation}. Spectroscopic studies of polluted white dwarfs enable the chemical compositions of the polluting bodies to be determined. These systems uniquely allow the measurement of the bulk compositions of extrasolar planetesimals \citep[e.g.][]{klein2011rocky, jura2012two}. To date, 20 different heavy elements have been detected in polluted white dwarfs; in a given star (GD 362) the most heavy elements detected is 19 \citep[e.g.][]{zuckerman2007chemical,xu2013two,melis2016does,xu2017chemical}. The abundances resemble solar system analogues, with most pollutants resembling bulk Earth to zeroth order. Some bodies also show evidence for differentiation, posing questions about collisional processing \citep[e.g.][]{hollands2018cool, harrison2018polluted}. 

In addition to atmospheric pollution, G29-38 was the first polluted white dwarf also discovered to show an infrared excess \citep{zuckerman1987excess}. Follow up observations confirmed its dusty nature \citep{graham1990infrared}. \textit{Spitzer} drastically increased the number of dust detections around polluted white dwarfs. To date, \textit{Spitzer} has confirmed that nearly 40 white dwarfs display an infrared excess from dust; this is calculated as between 1.5\% and 4\% of white dwarfs \citep[e.g.][]{becklin2005dusty, kilic2006debris, jura2007externally,rebassa2019infrared, wilson2019unbiased}. The temperature of the excess infrared emission is of the order 1000\,K \citep[e.g.][]{jura2007infrared}, demonstrating the dust lies close to the host star. Eight of the white dwarfs with dust also show evidence of circumstellar gas in emission near the same radius as the dust \citep{gaensicke2006gaseous, gansicke2007sdss, gansicke2008sdss, melis2010echoes, farihi2012trio, melis2012gaseous, brinkworth2012spitzer, debes2012detection}.  

Additional support for the asteroid tidal disruption model is the discovery of disintegrating planetesimals transiting the polluted and dusty white dwarf WD 1145+017 \citep{vanderburg2015disintegrating}. The deepest transit of this star has a period of 4.5 hours and blocks 60\% of the flux from the star in the optical \citep{rappaport2016drifting, gary2016wd}. Another white dwarf, ZTF J013906.17+524536.89, has also been discovered with transit features. The transits cause 30-45\% drops in the optical flux, and are separated by 110 days \citep{vanderbosch2019white}.

For the asteroid tidal disruption model, planetary bodies need to be scattered inwards from the surviving outer planetary system of the white dwarf. There are many dynamical mechanisms which can perturb planetesimals on to star-grazing orbits, the favoured theories cite planets as the source. Stellar mass loss can induce instabilities whereby a planet or planets perturb planetesimals on to star-grazing orbits \citep{debes2002there,bonsor2011dynamical,veras2011great,debes2012link, mustill2018unstable}. \citet{gansicke2019accretion} found evidence for the presence of a close-in planet around a white dwarf, demonstrating planets can survive to the white dwarf phase. Alternative mechanisms to scatter planetesimals include: perturbations due to wide binary companions \citep{hamers2016white,veras2016binary,petrovich2017planetary,stephan2017throwing,smallwood2018white}, although \citet{wilson2019unbiased} suggest these may be unimportant, and the liberation of exo-moons \citep{payne2016liberating}.

The scattering of planetary bodies that leads to the pollution is expected to be a stochastic process \citep{wyatt2014stochastic}, with variability predicted on human time-scales. Variations could come in the form of gaseous emission, dusty emission, or total flux caused by transiting debris (as mentioned previously); we may observe variability at optical and infrared wavelengths. These examples have all been seen in polluted white dwarf systems. Variability in the dust emission has been seen for a few white dwarfs. SDSS J095904.69$-$020047.6 (hereafter WD J0959$-$0200) and WD 1226+110 demonstrated large drops in dust flux \citep{xu2014drop, xu2018infrared}, whilst WD 0408$-$041 among other white dwarfs show both increases and decreases in the flux \citep{farihi2018dust, swan2019most, wang2019ongoing}. Variability in the strength of the 10$\micron$ dust feature has also be seen for G29-38 \citep{xu2018infrared}. Variability in the circumstellar gaseous material has been observed for a handful of white dwarfs \citep{wilson2014variable, wilson2015composition, manser2015doppler, manser2016another, dennihy2018rapid}, this is mostly attributed to the precession of eccentric rings, or planetesimals orbiting within the debris \citep{manser2019planetesimal}.

The first white dwarf discovered to show dust emission variability was WD J0959$-$0200. The 3--5\,$\micron$ flux dropped by 35\% between observations separated by a year and appeared to be stable afterwards, and the near-infrared \textit{K} band flux dropped by 18.5\% between observations separated by 8 years \citep{xu2014drop}. The large \textit{K} band excess for this white dwarf means the dust is hot and close in. These findings imply the variability event was large and became stable quickly. Motivated by the large variability in WD J0959$-$0200, we present a study to search for near-infrared variability in 34 white dwarfs with infrared excesses. We aim to characterise how white dwarfs vary in the near-infrared in order to better understand how dusty material arrives in the atmospheres of the white dwarfs.

This paper is structured as follows. Section \ref{UKIRT} describes the infrared monitoring campaign over three years for 34 white dwarfs with infrared excesses using the UK Infrared Telescope (UKIRT). Section \ref{Results} describes the variability analysis method and the results of this analysis. Section \ref{Discussion} discusses further interpretations from the UKIRT survey and the link with the mid-infrared. Section \ref{Conclusion} describes the main conclusions from the study.

\section{UKIRT Infrared Monitoring Campaign}  \label{UKIRT}

\begin{table*}
	\centering
	\caption{The UKIRT sample of dusty polluted white dwarfs and their properties.}
	\label{tab:WDs}
	\begin{tabular}{lllcccc} % four columns, alignment for each
		\hline
		WD Name & Other Name & Gaia DR2 Number & Composition$^{**}$ & T${_{eff} }^{\star}$ & log(g)$^{\star}$ & Nature of Excess $^{\dagger}$ \\
		\hline	
    WD 0010+280 & PG 0010+281 & 2859951106737135488 & H & 23700  & 7.73$^{\,1}$  & Dust$^{\,\alpha}$  \\
WD 0106$-$328 & HE 0106$-$3253 & 5026963661794939520 & H & 16200  & 8.03$^{\,1}$  & Dust$^{\,\beta}$ \\
WD 0146+187 & GD 16 & 95297185335797120 & He & 11500  & 8.00$^{\,2}$ & Dust$^{\,\gamma}$  \\
WD 0300$-$013 & GD 40 & 5187830356195791488 & He & 13600  & 8.02$^{\,3}$  & Dust$^{\,\delta}$  \\
WD 0307+077 & HS 0307+0746 & 13611477211053824 & H & 10100 & 7.99$^{\,1}$ &  Dust$^{\,\beta}$  \\
WD 0408$-$041 & GD 56 & 3251748915515143296 & H & 14200 & 7.96$^{\,1}$  & Dust$^{\,\delta}$ \\
WD 0435+410 & GD 61 & 203931163247581184 & He & 15700  & 8.04$^{\,3}$  & Dust$^{\,\epsilon}$  \\
WD J0738+1835$^{*}$ & SDSS J073842.57+183509.6 & 671450448046315520 & He & 14000 & 8.40$^{\,4}$ & Dust$^{\,\zeta}$  \\
WD J0959$-$0200$^{*}$ & SDSS J095904.69$-$020047.6 & 3829599892897720832 & H & 13300 & 8.06$^{\,5}$ & Dust$^{\,\eta}$  \\
WD 1015+161 & PG 1015+161 & 3888723386196630784 & H & 19200  & 8.03$^{\,1}$  & Dust$^{\,\delta}$ \\
WD 1018+410 & PG 1018+411 & 804169064957527552 & H & 21700  & 8.05$^{\,1}$  & Dust$^{\,\theta}$ \\
WD 1041+092 & ... & 3869060540584643328 & H & 17600  & 8.12$^{\,1}$  & Dust$^{\,\zeta}$ \\
WD 1116+026 & GD 133 & 3810933247769901696 & H & 12600  & 8.04$^{\,6}$  & Dust$^{\,\delta}$  \\
WD 1145+017 & HE 1145+0145 & 3796414192429498880 & He & 15900  & 8.00$^{\,7}$  & Dust$^{\,\iota}$ \\
WD 1145+288 & ... & 4019789359821201536 & H & 12400 & 7.96$^{\,1}$  & Cand$^{\,\kappa}$ \\
WD 1150$-$153 & EC 11507$-$1519 & 3571559292842744960 & H & 11400  & 7.98$^{\,1}$  & Dust$^{\,\lambda}$ \\
WD J1221+1245$^{*}$ & SDSS J122150.81+124513.3 & 3908649148232998400 & H & 12300  & 8.20$^{\,5}$ & Dust$^{\,\eta}$ \\
WD 1225$-$079 & PG 1225$-$079 & 3583181371265430656 & He & 10800  & 8.00$^{\,8}$  & Dust$^{\,\beta}$  \\
WD 1226+110 & ... & 3904415787947492096 & H & 20900  & 8.11$^{\,1}$  & Dust$^{\,\mu}$ \\
WD 1232+563 & ...  & 1571584539980588544 & He & 11800 & 8.30$^{\,3}$ & Cand$^{\,\kappa}$ \\
WD 1349$-$230 & HE 1349$-$2305 & 6287259310145684608 & He & 18200 & 8.13$^{\,9}$  & Dust$^{\,\nu}$  \\
WD 1456+298 & G166-58 & 1281989124439286912 & H & 7390  & 8.00$^{\,1}$  & Dust$^{\,\xi}$ \\
WD 1504+329 & ... & 1288812212565231232 & H & 7180 & 8.06$^{\,10}$ & Cand$^{\,\kappa}$ \\
WD 1536+520 & ... & 1595298501827000960 & H & 16700 & 7.70$^{\,6}$ & Cand$^{\,\kappa}$ \\
WD 1551+175 & ... & 1196531988354226560 & He & 15600 & 7.96$^{\,11}$ & Dust$^{\,\pi}$ \\
WD 1554+094 & KUV 15519+1730 & 4454599238843776128 & H &  21800 & 7.63$^{\,12}$ & BD+Dust$^{\,\rho}$ \\
WD J1617+1620$^{*}$ & SDSS J161717.04+162022.4 & 4465269178854732160 & H & 13200  & 8.04$^{\,1}$ & Dust$^{\,\zeta}$ \\
WD 1729+371 & GD 362 & 1336442472164656000 & H & 10300 & 8.13$^{\,6}$ & Dust$^{\,\xi}$  \\
WD 1929+011 & ... & 4287654959563143168 & H & 25400 & 8.17$^{\,1}$  & Dust$^{\,\theta}$ \\
WD 2132+096 & HS 2132+0941 & 1742342784582615936 & H & 13100  & 7.96$^{\,13}$  & Dust$^{\,\pi}$ \\
WD 2207+121 & ... & 2727904257071365760 & He & 14800  & 7.97$^{\,3}$ & Dust$^{\,\sigma}$ \\
WD 2221$-$165 & HE 2221$-$1630 & 2595728287804350720 & H & 9900  & 8.12$^{\,1}$  & Dust$^{\,\beta}$ \\
WD 2326+049 & G29-38 & 2660358032257156736 & H & 11200  & 8.00$^{\,1}$ & Dust$^{\,\tau}$  \\
WD 2328+107 & PG 2328+108 &  2762605088857836288 & H & 17900  & 7.74$^{\,1}$ & Dust$^{\,\theta}$ \\

		\hline
	\end{tabular}
	\begin{tablenotes}
      \small
      \item Notes: 
      \item $^{*}$ No WD name, so an SDSS nickname is used.
      \item $^{**}$ Dominant element in the atmosphere.
      \item $^{\star}$ The temperature and log(g) come from photometric fits, where possible. For a number of the white dwarfs, if no photometric fit was available, we performed a new  fit considering the \textit{Gaia} distances. See Section \ref{Phot_Fits} for further information.
      
      \item $^{\star}$ Properties of the sample references: (1) This work, (2) \citet{koester2005hs}, (3) \citet{coutu2019analysis}, (4) \citet{dufour2012detailed}, (5) \citet{farihi2012trio}, (6) \citet{dufour2017mwdd}, (7) \citet{vanderburg2015disintegrating}, (8) \citet{klein2011rocky}, (9) \citet{voss2007high}, (10) \citet{barber2014dusty}, (11) \citet{bergeron2011comprehensive}, (12)   \citet{farihi2017circumbinary}, (13) \citet{gentile2018gaia}.
      
      \item $^{\dagger}$ For the nature of the infrared excess: if the white dwarf has a dust detection confirmed with \textit{Spitzer} observations, it is labelled `dust' and the reference is that of the \textit{Spitzer} observations, otherwise it is a candidate, and it is labelled `cand', so the reference highlights the paper at which the dust emission is first referred. WD 1554+094 has a brown dwarf companion and dust.
      \item $^{\dagger}$ Nature of infrared excess references: ($\alpha$) \citet{xu2015young}, ($\beta$) \citet{farihi2010strengthening}, ($\gamma$) \citet{farihi2009infrared}, ($\delta$) \citet{jura2007externally}, ($\epsilon$) \citet{farihi2011possible}, ($\zeta$) \citet{brinkworth2012spitzer}, ($\eta$) \citet{farihi2012trio}, ($\theta$) \citet{rocchetto2015frequency}, ($\iota$) \citet{xu2017dearth}, ($\kappa$)  \citet{debes2011wired}, ($\lambda$) \citet{jura2009six}, ($\mu$) \citet{brinkworth2009dusty}, ($\nu$) \citet{girven2012constraints}, ($\xi$) \citet{farihi2008spitzer}, ($\pi$) \citet{bergfors2014signs}, ($\rho$) \citet{farihi2017circumbinary}, ($\sigma$) \citet{xu2012spitzer}, ($\tau$) \citet{reach2005dust}.

    \end{tablenotes}
\end{table*}

An infrared monitoring campaign was completed using the Wide Field Camera (WFCAM) on the UKIRT on Maunakea, Hawaii. WFCAM operates in the near-infrared between 0.83 and 2.37\,$\micron$; this includes the \textit{J}, \textit{H} and \textit{K} filters \citep{hewett2006ukirt}. WFCAM has four detectors composed of 2048 by 2048 18 micron pixels, with a pixel scale of 0.4$''$. 

Our sample consists of 34 white dwarfs with infrared excesses. Table \ref{tab:WDs} shows our sample, giving the white dwarf parameters and the nature of the infrared excess. In our sample, 30 white dwarfs have dust emission confirmed with \textit{Spitzer} observations, the 4 remaining objects are identified by WISE and are labelled as dust candidates. \textit{JHK} broadband photometry was obtained for the 34 targets with the UKIRT. 48 hours of UKIRT WFCAM time between semesters 2014B and 2017A were used to study long-term variability in the near-infrared. The dates of the observations for the 34 white dwarfs are shown in Table \ref{tab:DATES}.

Each white dwarf was observed up to 6 times across the 3 year campaign. Our observations probed short time-scales (minutes) between individual frames taken on a given night, and long time-scales (years) between observation dates. Further discussions about time-scales are given in Section \ref{timescales-section}. The observations were designed such that a signal-to-noise ratio (SNR) of 30 was obtained for the \textit{J} and \textit{H} bands, and 40 for the \textit{K} band over the set of observations for that filter. To achieve the required SNR, the brightest white dwarfs in the sample required 25\,s of exposure time for the \textit{J} and \textit{H} band, and 75\,s for the \textit{K} band. For the faintest white dwarfs, this required 150\,s for the \textit{J} band, 450\,s for the \textit{H} band, and 1250\,s for the \textit{K} band. As the infrared sky is bright, each frame consisted of a dithered stack of five 5\,s or 10\,s exposures. Table \ref{tab:Frames} shows the exposure time for each of the five individual exposures making up the dithered stack. All data obtained with WFCAM were pipeline-processed by the Cambridge Astronomical Survey Unit using standard infrared photometry data reduction steps \citep[CASU,][]{irwin2004vista,dye2006ukirt}. 

Further processing using the \textsc{lightcurves} software\footnote{https://github.com/mdwarfgeek/lightcurves} was executed to improve the precision of the photometry  \citep{irwin2007monitor}. List driven photometry was performed using a master frame to force the centroid positions for the objects. Here, the master frame is the stacked image of all frames; this increases the SNR of the master frame and reduces errors associated with inaccurately placed centroids. Each frame used in the construction of the lightcurves is re-aligned to the same set of astrometric calibrators (from 2MASS) as the master frame, which gives significantly improved precision in aperture location when compared to centroiding for faint sources \citep[see][]{irwin2007monitor}. For a particular object, all frames had an aperture between 3--5 pixels (1.2--2$''$), depending on which aperture gave the lowest RMS value. Using a large number of non-variable stars, the zero-point shift for each frame was calculated to reduce atmospheric effects, and a 2D polynomial was fitted to the magnitude residuals to minimise the error associated with the position on the detector. \textsc{lightcurves} outputted a robustly calibrated light curve for each filter and for each object in the catalogue list. For variability searches this is essential as it allows us to probe changes down to the percent level. 

\section{Variability Analysis} \label{Results}

\subsection{Analysis Method} \label{Method}

\begin{figure}
   \centering 
   \includegraphics[width=0.48\textwidth]{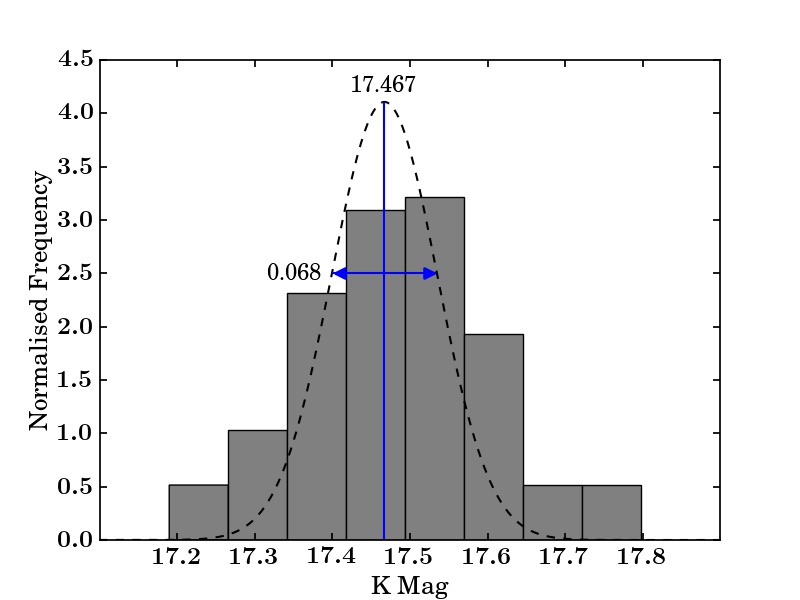}
   \caption{A normalised histogram showing the distribution of photometrically corrected magnitudes from the \textsc{lightcurves} software using all dithered stacked frames for WD 1145+017. Each dithered stacked frame contains $5 \times 10$ second exposures. This white dwarf has 102 measurements in the \textit{K} band over 782 days. The Gaussian curve shows the median best-fitting Gaussian, with a median magnitude of 17.467, and a median standard deviation of 0.068 mags.} 
  
  \label{fig:Gaussians}
\end{figure}

This work aimed to obtain high precision \textit{JHK} photometry of a sample of dusty white dwarfs and to search for and constrain the level of variability in the photometry over the length of the survey. Throughout the analysis the Vega magnitude system was adopted. In order to robustly study the statistics of the photometry, the same analysis method was applied to all white dwarf observations regardless of the magnitude of the white dwarf and the number of measurements in the \textit{J}, \textit{H} and \textit{K} bands. Gaussians were fitted to the distribution of photometrically corrected magnitudes from the \textsc{lightcurves} software using all dithered stacked frames, as demonstrated in Fig. \ref{fig:Gaussians}. The total number of dithered stacked frames for each filter is shown in Table \ref{tab:Frames}, this ranges from 3 to 125, with the median number of frames for the \textit{J}, \textit{H} and \textit{K} bands being 9, 12 and 20 respectively. A Markov Chain Monte Carlo (MCMC) approach was implemented to model the magnitude distribution of each star with a Gaussian profile, yielding posterior distributions for the mean and standard deviation thereof. In our analysis, we adopted the median values from the posteriors. Uniform priors were used for the magnitude between 0 and 20, and standard deviation between 0 and 2. We tested larger upper values for the assumed prior range to test the sensitivity on the parameters. The results remained consistent independent of the size of the prior. A python package, \textsc{pymc} \citep{patil2010pymc}, was used for MCMC parameter estimation. Five walkers were used, each resulting in 100,000 posterior samples, with the first 40\% of the chain being discarded in the burn-in phase. The Gelman-Rubin statistic \citep{gelman1992inference} was implemented to check for chain convergence. For the two parameters (magnitude and standard deviation), all white dwarfs had values 1.000\,$<$\,GR\,$<$\,1.037 for the \textit{J} band, 0.999\,$<$\,GR\,$<$\,1.002 for the \textit{H} band, and  1.000\,$<$\,GR\,$<$\,1.005 for the \textit{K} band, demonstrating good convergence. The median value of the posterior distribution of the standard deviation was used for the variability measurement. The errors are quoted using the 16th and 84th percentiles of the posterior distribution. This median variability measurement is a robust and reliable way of measuring variability.

To ensure that any variability detected was real, stellar objects in the field of view were analysed using the same approach for comparison. For all field stars the median value of the standard deviation of the Gaussian as a function of magnitude was used to represent the sensitivity of the survey as a function of magnitude. Using the median ensures no variable field stars are contaminating the function. The white dwarfs were then compared to the median MCMC standard deviation of the non-variable field stars at the same magnitude as the white dwarf. If the standard deviation of the white dwarf was significantly higher than that of the field stars, then they were identified as variable. The field objects selected for the comparison were those classified by \textsc{lightcurves} as stellar and those observed in the same detector as the white dwarf. The number of field stars  used in the comparison for each white dwarf ranged from 110 for white dwarfs in sparse fields of view up to 6700 for white dwarfs in densely populated regions. The median number of field stellar objects used in the comparison for each white dwarf was 210. 

\subsection{Variability Results}
{
\renewcommand{\arraystretch}{1.4}
\begin{table*}
	\centering
	\caption{The resulting median magnitude and standard deviation found from the Gaussian fits to the photometry from the UKIRT observations in the \textit{J}, \textit{H} and \textit{K} bands. The errors are quoted as the 16th and 84th percentiles from the posterior distributions.}
	\label{tab:Var}
	\begin{tabular}{lcccccc}
		\hline
		WD Name & J Mag & J Std (Mags) & H Mag & H Std (Mags) & K Mag & K Std (Mags) \\
		\hline
WD 0010+280& $16.304_{-0.014}^{+0.014}$ & $0.016_{-0.011}^{+0.019} $ & $16.375_{-0.011}^{+0.011}$ & $0.014_{-0.010}^{+0.016} $ & $16.345_{-0.018}^{+0.019}$ & $0.043_{-0.019}^{+0.023} $ \\
WD 0106$-$328& $15.616_{-0.009}^{+0.009}$ & $0.008_{-0.006}^{+0.011} $ & $15.831_{-0.020}^{+0.020}$ & $0.041_{-0.017}^{+0.026} $ & $15.836_{-0.034}^{+0.035}$ & $0.038_{-0.027}^{+0.049} $ \\
WD 0146+187& $15.679_{-0.007}^{+0.007}$ & $0.006_{-0.004}^{+0.008} $ & $15.650_{-0.010}^{+0.010}$ & $0.013_{-0.009}^{+0.014} $ & $15.472_{-0.029}^{+0.029}$ & $0.060_{-0.025}^{+0.039} $ \\
WD 0300$-$013& $15.888_{-0.006}^{+0.006}$ & $0.006_{-0.004}^{+0.007} $ & $15.873_{-0.013}^{+0.013}$ & $0.017_{-0.012}^{+0.017} $ & $15.861_{-0.014}^{+0.014}$ & $0.022_{-0.015}^{+0.020} $ \\
WD 0307+077& $16.274_{-0.013}^{+0.013}$ & $0.012_{-0.008}^{+0.016} $ & $16.159_{-0.011}^{+0.012}$ & $0.016_{-0.011}^{+0.017} $ & $16.217_{-0.018}^{+0.018}$ & $0.046_{-0.015}^{+0.020} $ \\
WD 0408$-$041& $16.005_{-0.006}^{+0.006}$ & $0.007_{-0.005}^{+0.008} $ & $15.909_{-0.012}^{+0.012}$ & $0.010_{-0.007}^{+0.012} $ & $14.968_{-0.011}^{+0.011}$ & $0.012_{-0.008}^{+0.014} $ \\
WD 0435+410& $15.305_{-0.018}^{+0.017}$ & $0.022_{-0.015}^{+0.055} $ & $15.318_{-0.037}^{+0.043}$ & $0.055_{-0.035}^{+0.138} $ & $15.271_{-0.011}^{+0.011}$ & $0.010_{-0.007}^{+0.012} $ \\
WD J0738+1835& $18.080_{-0.016}^{+0.015}$ & $0.026_{-0.017}^{+0.022} $ & $17.857_{-0.012}^{+0.012}$ & $0.028_{-0.018}^{+0.019} $ & $17.329_{-0.011}^{+0.011}$ & $0.065_{-0.012}^{+0.012} $ \\
WD J0959$-$0200& $18.388_{-0.025}^{+0.027}$ & $0.044_{-0.029}^{+0.037} $ & $18.221_{-0.021}^{+0.022}$ & $0.090_{-0.021}^{+0.024} $ & $17.809_{-0.015}^{+0.015}$ & $0.085_{-0.019}^{+0.019} $ \\

WD 1015+161& $16.044_{-0.012}^{+0.013}$ & $0.016_{-0.011}^{+0.020} $ & $16.114_{-0.015}^{+0.016}$ & $0.019_{-0.013}^{+0.023} $ & $16.016_{-0.018}^{+0.018}$ & $0.018_{-0.012}^{+0.022} $ \\
WD 1018+410& $16.954_{-0.022}^{+0.024}$ & $0.029_{-0.021}^{+0.039} $ & $16.964_{-0.024}^{+0.025}$ & $0.045_{-0.024}^{+0.037} $ & $16.835_{-0.019}^{+0.019}$ & $0.026_{-0.018}^{+0.026} $ \\
WD 1041+092& $17.701_{-0.014}^{+0.014}$ & $0.014_{-0.010}^{+0.016} $ & $17.690_{-0.015}^{+0.015}$ & $0.020_{-0.014}^{+0.021} $ & $17.761_{-0.018}^{+0.018}$ & $0.037_{-0.024}^{+0.028} $ \\
WD 1116+026& $14.852_{-0.008}^{+0.008}$ & $0.008_{-0.006}^{+0.013} $ & $14.825_{-0.013}^{+0.013}$ & $0.013_{-0.009}^{+0.023} $ & $14.795_{-0.007}^{+0.006}$ & $0.005_{-0.004}^{+0.006} $ \\
WD 1145+017& $17.671_{-0.031}^{+0.031}$ & $0.118_{-0.024}^{+0.030} $ & $17.680_{-0.015}^{+0.015}$ & $0.075_{-0.014}^{+0.016} $ & $17.467_{-0.012}^{+0.012}$ & $0.068_{-0.015}^{+0.015} $ \\
WD 1145+288& $17.671_{-0.016}^{+0.016}$ & $0.015_{-0.011}^{+0.017} $ & $17.664_{-0.015}^{+0.015}$ & $0.040_{-0.022}^{+0.020} $ & $17.281_{-0.012}^{+0.012}$ & $0.056_{-0.017}^{+0.016} $ \\
WD 1150$-$153& $16.236_{-0.007}^{+0.007}$ & $0.008_{-0.006}^{+0.009} $ & $16.189_{-0.010}^{+0.010}$ & $0.011_{-0.008}^{+0.012} $ & $15.847_{-0.011}^{+0.011}$ & $0.014_{-0.010}^{+0.014} $ \\
WD J1221+1245& $18.413_{-0.029}^{+0.029}$ & $0.039_{-0.027}^{+0.037} $ & $18.264_{-0.024}^{+0.026}$ & $0.103_{-0.027}^{+0.029} $ & $17.841_{-0.013}^{+0.013}$ & $0.039_{-0.025}^{+0.026} $ \\
WD 1225$-$079& $14.963_{-0.008}^{+0.009}$ & $0.010_{-0.007}^{+0.014} $ & $14.960_{-0.015}^{+0.015}$ & $0.013_{-0.010}^{+0.023} $ & $14.957_{-0.007}^{+0.007}$ & $0.012_{-0.008}^{+0.010} $ \\
WD 1226+110& $16.965_{-0.012}^{+0.012}$ & $0.011_{-0.008}^{+0.013} $ & $16.944_{-0.011}^{+0.011}$ & $0.012_{-0.008}^{+0.013} $ & $16.623_{-0.015}^{+0.015}$ & $0.029_{-0.019}^{+0.022} $ \\
WD J1234+5606& $18.279_{-0.023}^{+0.023}$ & $0.035_{-0.023}^{+0.032} $ & $18.240_{-0.022}^{+0.023}$ & $0.056_{-0.033}^{+0.032} $ & $17.797_{-0.020}^{+0.021}$ & $0.122_{-0.020}^{+0.022} $ \\
WD 1349$-$230& $17.018_{-0.011}^{+0.011}$ & $0.016_{-0.011}^{+0.015} $ & $17.056_{-0.013}^{+0.013}$ & $0.022_{-0.015}^{+0.019} $ & $16.982_{-0.017}^{+0.018}$ & $0.031_{-0.021}^{+0.027} $ \\
WD 1456+298& $14.995_{-0.012}^{+0.013}$ & $0.020_{-0.012}^{+0.025} $ & $15.062_{-0.012}^{+0.012}$ & $0.012_{-0.009}^{+0.021} $ & $14.830_{-0.009}^{+0.009}$ & $0.014_{-0.010}^{+0.013} $ \\
WD 1504+329& $18.234_{-0.036}^{+0.038}$ & $0.061_{-0.041}^{+0.055} $ & $18.056_{-0.017}^{+0.018}$ & $0.052_{-0.028}^{+0.026} $ & $18.029_{-0.021}^{+0.022}$ & $0.142_{-0.022}^{+0.023} $ \\
WD 1536+520& $17.749_{-0.011}^{+0.011}$ & $0.011_{-0.008}^{+0.012} $ & $17.546_{-0.011}^{+0.011}$ & $0.019_{-0.013}^{+0.016} $ & $16.830_{-0.008}^{+0.009}$ & $0.042_{-0.010}^{+0.010} $ \\
WD 1551+175 & $17.642_{-0.015}^{+0.015}$ & $0.028_{-0.017}^{+0.021} $ & $17.736_{-0.013}^{+0.014}$ & $0.031_{-0.019}^{+0.019} $ & $17.708_{-0.017}^{+0.018}$ & $0.086_{-0.021}^{+0.021} $ \\
WD 1554+094& $19.019_{-0.033}^{+0.033}$ & $0.039_{-0.027}^{+0.042} $ & $18.973_{-0.042}^{+0.045}$ & $0.193_{-0.042}^{+0.047} $ & $18.662_{-0.034}^{+0.035}$ &  $ 0.194_{-0.036}^{+0.037}$ \\
WD J1617+1620& $17.337_{-0.011}^{+0.011}$ & $0.018_{-0.012}^{+0.016} $ & $17.202_{-0.013}^{+0.013}$ & $0.049_{-0.013}^{+0.015} $ & $16.865_{-0.012}^{+0.012}$ & $0.026_{-0.017}^{+0.020} $ \\
WD 1729+371& $16.130_{-0.005}^{+0.005}$ & $0.008_{-0.005}^{+0.007} $ & $16.068_{-0.007}^{+0.007}$ & $0.012_{-0.008}^{+0.011} $ & $15.853_{-0.006}^{+0.006}$ & $0.008_{-0.005}^{+0.008} $ \\
WD 1929+011& $14.820_{-0.006}^{+0.006}$ & $0.006_{-0.004}^{+0.008} $ & $14.843_{-0.007}^{+0.007}$ & $0.011_{-0.007}^{+0.009} $ & $14.656_{-0.007}^{+0.007}$ & $0.007_{-0.005}^{+0.008} $ \\
WD 2132+096& $16.238_{-0.005}^{+0.005}$ & $0.006_{-0.004}^{+0.006} $ & $16.283_{-0.009}^{+0.009}$ & $0.015_{-0.010}^{+0.013} $ & $16.282_{-0.008}^{+0.008}$ & $0.017_{-0.011}^{+0.012} $ \\
WD 2207+121& $17.600_{-0.011}^{+0.011}$ & $0.030_{-0.013}^{+0.014} $ & $17.452_{-0.010}^{+0.011}$ & $0.037_{-0.014}^{+0.013} $ & $17.245_{-0.011}^{+0.012}$ & $0.062_{-0.013}^{+0.013} $ \\
WD 2221$-$165& $15.978_{-0.010}^{+0.009}$ & $0.011_{-0.008}^{+0.012} $ & $15.902_{-0.009}^{+0.009}$ & $0.010_{-0.007}^{+0.011} $ & $15.858_{-0.021}^{+0.020}$ & $0.033_{-0.022}^{+0.029} $ \\
WD 2326+049& $13.143_{-0.078}^{+0.080}$ & $0.113_{-0.058}^{+0.229} $ & $13.038_{-0.049}^{+0.053}$ & $0.071_{-0.037}^{+0.178} $ & $12.652_{-0.006}^{+0.006}$ & $0.011_{-0.007}^{+0.009} $ \\
WD 2328+107& $15.999_{-0.004}^{+0.004}$ & $0.007_{-0.005}^{+0.006} $ & $16.004_{-0.006}^{+0.006}$ & $0.006_{-0.004}^{+0.007} $ & $16.071_{-0.011}^{+0.011}$ & $0.024_{-0.015}^{+0.015} $ \\

		\hline
	\end{tabular}
\end{table*}
}

Table \ref{tab:Var} shows the results of the MCMC runs, and lists the median magnitude and median standard deviation on this magnitude for all white dwarfs in the survey. Fig. \ref{fig:MCMC_plots} plots these median magnitudes and standard deviations for the white dwarfs and field stars in the \textit{J}, \textit{H} and \textit{K} bands. Most white dwarfs have standard deviations which are consistent with that of the field stellar objects. This implies that most of the white dwarfs have \textit{J}, \textit{H} and \textit{K} fluxes which are no more variable than the field stars, which we take to be a proxy of the sensitivity of the survey as a function of magnitude.

A typical spectral energy distribution (SED) for a white dwarf that is representative of our sample peaks in the optical, whereas the dust emission peaks in the infrared. Of the near-infrared bands, the \textit{K} band is more sensitive to the dust emission. Consistent fluxes in the \textit{K} band imply there was little change in the \textit{K} band dust emission for the white dwarfs. A few objects stand out from the plots, these objects are discussed further in the following section. We also comment on some previously variable objects from the literature that were observed in this monitoring campaign.

\begin{figure*}
\centering
\subfloat[\textit{J} Band Observations]{
  \includegraphics[width=0.85\textwidth]{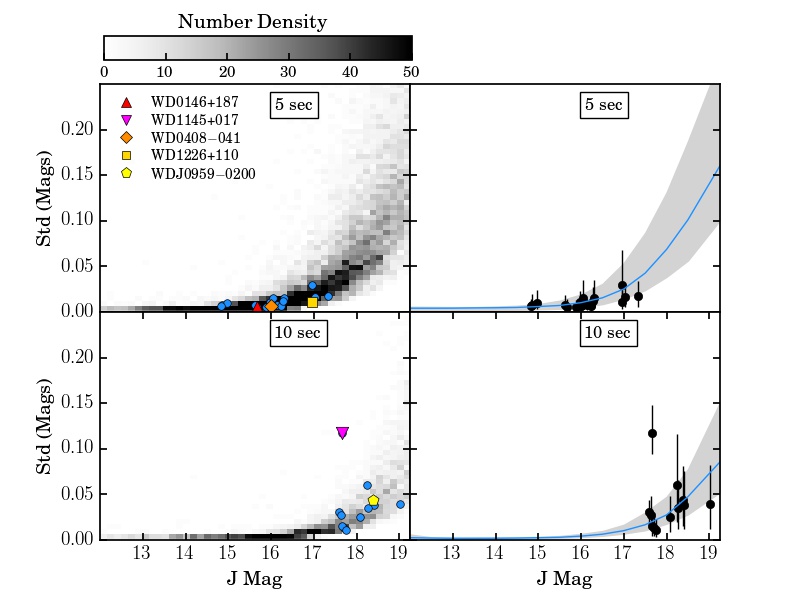}
  \label{fig:MCMC_plots_J}
}
\vspace{0.1cm}
\subfloat[\textit{H} Band Observations]{
  \includegraphics[width=0.85\textwidth]{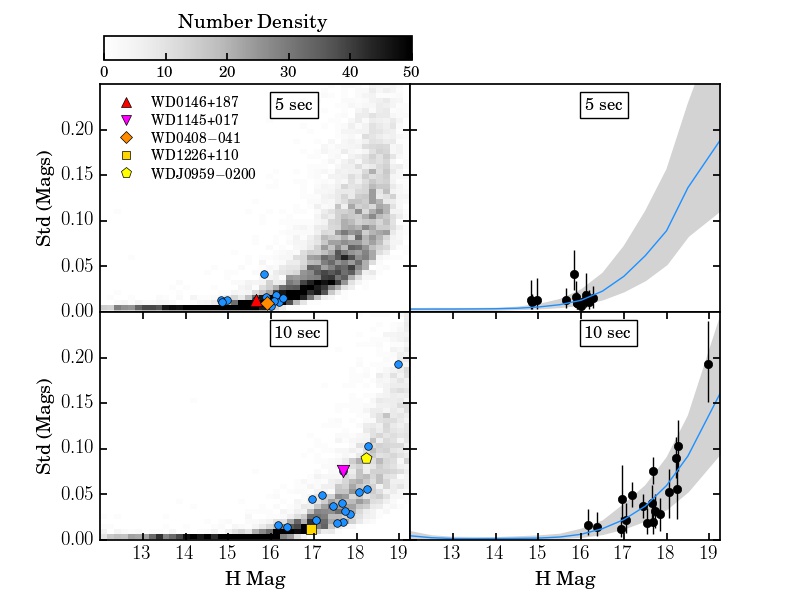}
   \label{fig:MCMC_plots_H}
}
\vspace{0.1cm}
\caption{ }
\end{figure*}

\begin{figure*}
\ContinuedFloat
\subfloat[\textit{K} Band Observations]{
  \includegraphics[width=0.85\textwidth]{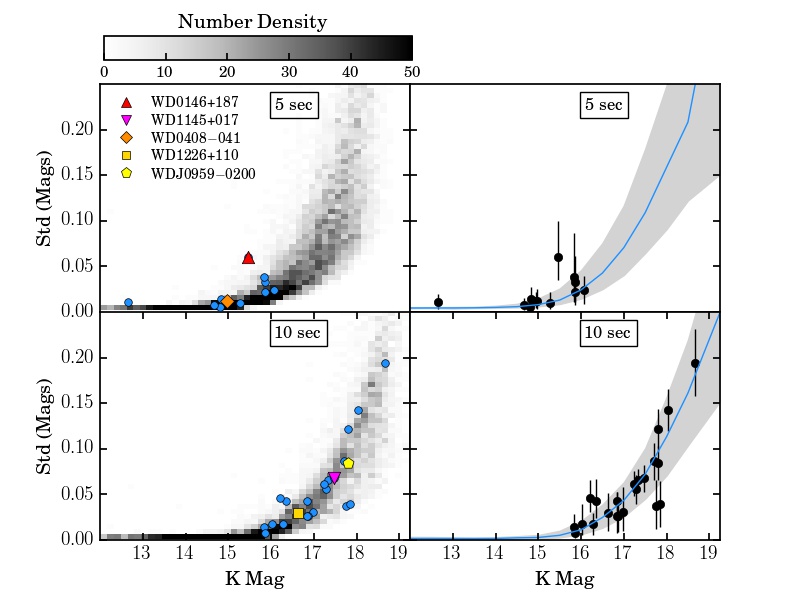}
    \label{fig:MCMC_plots_K}

}
\caption{The median observed magnitudes in the (a) \textit{J}, (b) \textit{H} and (c) \textit{K} bands plotted against the median standard deviation of this magnitude for all white dwarfs and field stars. The two different observing modes, with 5 and 10 second frame times, are analysed separately. The density plots on the left hand side show the parameters for all field stellar objects in the field of view, there are of the order $10^4$ field stars in each plot. The blue points show the white dwarfs, with the objects discussed in Section \ref{sources} labelled. The right hand plots show the median magnitude and standard deviation with the 16th and 84th percentile confidence levels for field stars in bins of width 0.5 mag. This forms a distribution about that median which should not comprise contaminant variable stars. The white dwarfs are plotted in black with the same values as in the left hand plots, but also including the 16th and 84th percentiles. This plot can be used to distinguish white dwarfs which are variable from those that follow the field star distribution.}
\label{fig:MCMC_plots}
\end{figure*}

\subsection{Discussion of Individual Sources:} \label{sources}

\subsubsection{WD 1145+017}

\citet{vanderburg2015disintegrating} discovered that WD 1145+017 is transited by multiple disintegrating planetesimals. The transit periods were approximately 4.5 hours and lasted various lengths from minutes to hours, with varying transit depths, the deepest of which blocks 60\% of the flux from the star in the optical. The transits were not always present, and the transit activity level was variable \citep{gary2016wd}.

\begin{figure*}
   \centering 
   \captionsetup{font=footnotesize}
   \includegraphics[width=1.0\textwidth]{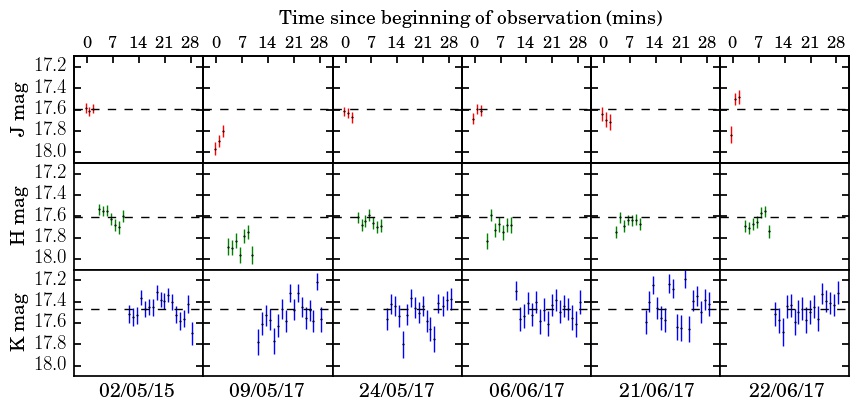}
    \caption{A light curve in the \textit{J}, \textit{H} and \textit{K} bands for WD 1145+017. At each observation date a number of observations were taken in the \textit{J}, \textit{H} and \textit{K} bands, totalling 28 minutes. The dotted lines show the mean magnitude for each band calculated from the first observation date, 02/05/15. This highlights how there was an apparent dip on 09/05/19 in the flux in the \textit{J} and \textit{H} bands attributed to dust transit features. }
    \label{fig:HE1145_LC}
\end{figure*}

The pink inverted triangle in Figs. \ref{fig:MCMC_plots_J}, \ref{fig:MCMC_plots_H}, and \ref{fig:MCMC_plots_K} show WD 1145+017 was variable at 4.1\,$\sigma$ in the \textit{J} band, 2.1\,$\sigma$ in the \textit{H} band, and was not significantly variable in the \textit{K} band. From the UKIRT light curves of this white dwarf, shown in Fig. \ref{fig:HE1145_LC}, it appears that the \textit{J} band and possibly \textit{H} band magnitudes were fainter on 09/05/2017. The \textit{J}, \textit{H} and \textit{K} band observations were taken consecutively, the length of the observations were 3.5, 7.5 and 17 minutes respectively. Therefore, as the median dip duration is 9 minutes \citep{gary2016wd}, it is possible that the transit occurred during the \textit{J} band observations, with an overlap into the \textit{H} band observations. Alternatively, as the variability was strongest in the \textit{J} band and weakest in the \textit{K} band it is consistent with the transit depth measurements found in \citet{xu2017dearth}, and could also explain the differences in the observed variability. Near contemporaneous optical observations indeed show transits in the light curve \citep[B. Gary, private communications\footnote{http://www.brucegary.net/zombie6/}][]{rappaport2017wd}. The decrease in the \textit{J} and \textit{H} band flux is very likely to be caused by transits, rather than dust variability.

\subsubsection{WD 0146+187}

The object with the most significant deviation in the \textit{K} band was WD 0146+187. In Fig. \ref{fig:MCMC_plots_K} the red triangle, which represents this object, shows a 2\,$\sigma$ variability in the \textit{K} band. There was a flux brightening of 0.14 mag from the 1st to the 2nd stacked frames on 03/10/14. There were no significant variations between the subsequent frames. Further investigations revealed there was nothing anomalous in the pixel map, and objects in the field of view within 100$''$ do not show the same brightening trend. This could be a real change, but as there is only one data point, and the variation is not statistically significant, it is difficult to make any substantial conclusions.

\subsubsection{WD J0959$-$0200}

WD J0959$-$0200 was previously observed to display a 35\% flux drop in 2010 in both \textit{Spitzer} 3.6 and 4.5\,$\micron$ Infrared Array Camera (IRAC) channels. The \textit{K} band flux also dropped by 18.5\% between 2005 and 2014 \citep{xu2014drop}. This variability measurement uses the peak-to-peak method, where the maximum flux change between two frames is quoted. After the \citet{xu2014drop} UKIRT data taken on the 27/03/14, in this work the \textit{K} band data were consistent between multiple observations spanning 1153 days (Nov 2014, Feb 2015, Mar 2016, May 2016, May 2017). The \textit{K} band flux was stable within 7.5\,$^{+1.6}_{-1.6}$\,\% ($\sigma =  0.085\,^{+0.019}_{-0.019}$) over these observations, seen in Fig. \ref{fig:MCMC_plots_K} as the yellow pentagon.

\subsubsection{WD 1226+110}

WD 1226+110 was found with a peak-to-peak drop in flux of 20\% in the \textit{Spitzer} 3.6 and 4.5\,$\micron$ bands, and 13\% in the \textit{K} band, when comparing the 2007 UKIRT Infrared Deep Sky Survey \textit{K} band flux with the 2015 UKIRT observation in this work \citep{xu2018infrared}. Since 2015, this UKIRT survey has shown no significant changes in flux for WD 1226+110 between multiple observations separated by 738 days (May 2015, Feb 2016, May 2016, May 2017). The \textit{K} band flux was stable within 2.7\,$^{+1.9}_{-1.7}$\,\% ($\sigma = 0.029\,^{+0.022}_{-0.019}$) for these observations, seen in Fig. \ref{fig:MCMC_plots_K} as the gold square. 

\subsubsection{WD 0408$-$041}

WD 0408$-$041 (GD56) has previously been observed with dust emission variations at 3--5\,$\micron$ showing brightening and dimming with peak-to-peak changes of 20\% over 11.2\,yrs \citep{farihi2018dust}. This was attributed to dust production and consequentially depletion. In this work, the fluxes were found to be consistent between multiple observations separated by 851 days (Oct 2014, Oct 2016, Jan 2017). The \textit{K} band dust emission appears to be stable within 1.1\,$^{+1.2}_{-0.8}$\,\% ($\sigma =  0.012\,^{+0.014}_{-0.008}$) for these observations, seen in Fig. \ref{fig:MCMC_plots_K} as the orange diamond. WD 0408$-$041 will be discussed again in Section \ref{Spitzer} in reference to the \textit{Spitzer} variability.

\subsection{Expected Variability}

{
\renewcommand{\arraystretch}{1.4}
\begin{table}
	\centering
	\caption{The level of variability that can be detected at each magnitude in the \textit{K} band, calculated from the field stars. The objects which were observed with 5 and 10 second frames were analysed separately. The median percentage variability is quoted for the field stars in bins with a width of one mag, along with the 16th and 84th percentiles.}
	\label{tab:Exp_Var}
	\begin{tabular}{ccc}
		\hline
		Magnitude & 5\,s variability (\%) & 10\,s variability (\%) \\
		\hline
11.0-12.0 & $0.41_{-0.02}^{+0.08}$ & $0.20_{-0.08}^{+0.27}$ \\
12.0-13.0 & $0.41_{-0.02}^{+0.07}$ & $0.15_{-0.04}^{+0.14}$ \\
13.0-14.0 & $0.43_{-0.04}^{+0.10}$ & $0.14_{-0.04}^{+0.12}$ \\
14.0-15.0 & $0.57_{-0.11}^{+0.29}$ & $0.20_{-0.07}^{+0.18}$ \\
15.0-16.0 & $1.27_{-0.51}^{+1.31}$ & $0.52_{-0.27}^{+0.58}$ \\
16.0-17.0 & $3.91_{-1.75}^{+3.47}$ & $2.32_{-1.17}^{+1.65}$ \\
17.0-18.0 & $9.43_{-3.89}^{+6.23}$ & $6.34_{-2.52}^{+3.33}$ \\
18.0-19.0 & $16.14_{-6.48}^{+7.67}$ & $13.16_{-4.62}^{+5.38}$ \\
		\hline
	\end{tabular}
\end{table}
}

We used the UKIRT survey to determine the level at which variability could be detected and ruled out. Table \ref{tab:Exp_Var} shows the level of variability that can be detected at each \textit{K} band magnitude. For each non-variable star in the fields of views of the white dwarfs, we have the median standard deviation of the magnitude measurements from the MCMC model. The level of variability is the median value of the non-variable stars' standard deviations, found in magnitude bins with a width of one mag. This median standard deviation was then converted to a percentage in flux space. We can rule out \textit{K} band flux variability at 1.3$_{-0.5}^{+1.3}$\,\% for objects brighter than 16th mag in the \textit{K} band, and at 9.4$_{-3.4}^{+6.2}$\,\% for objects brighter than 18th mag. 

\begin{figure}
   \centering 
   \captionsetup{font=footnotesize}
   \includegraphics[width=0.48\textwidth]{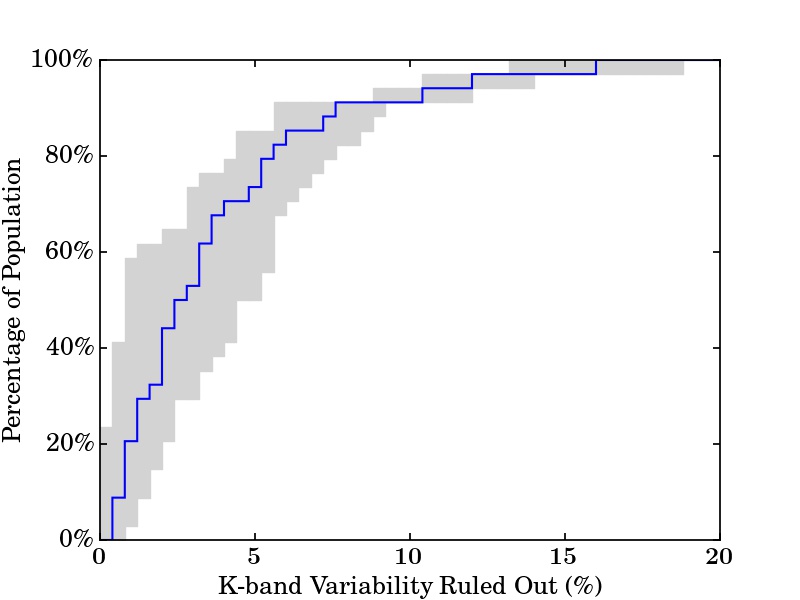}
   \caption{A cumulative distribution showing the level of \textit{K} band variability that can be ruled out for the white dwarfs. The blue line shows the median cumulative distribution, with the contours as the cumulative distribution on the upper and lower limits (16th and 84th percentiles). For 90\% of the sample, the \textit{K} band flux varied by less than 10\%.} 
  
  \label{fig:Percentage_Plot}
\end{figure}

Fig. \ref{fig:Percentage_Plot} shows a cumulative distribution of the level of variability we rule out for all white dwarfs. In other words, the percentage of white dwarfs where the \textit{K} band flux was seen to vary by no more than the plotted value. This was calculated as the median standard deviation converted to a percentage in flux space. This was also calculated for the 16th and 84th percentile errors, which is shown as the shaded region in the figure. The \textit{K} band flux of each white dwarf cannot have varied more than its equivalent standard deviation. Therefore, the \textit{K} band flux has varied at less than 16.4\,$\pm$\,2.8\% for the entire sample, and less than 2.88$^{+2.5}_{-1.9}$\% for the median case. For greater than 90\% of our sample we rule out variability at the 10\% level, and for the 10 most constrained cases, we rule out variability at the percent level, 1.57$^{+1.1}_{-1.0}$\%. 

Fig. \ref{fig:Timescale_Plot_2} shows the percentage variability we rule out for each white dwarf plotted against the maximum time-scale over which that white dwarf was monitored in this survey. For example, for the white dwarf monitored over the longest period, WD 1929+011, we rule out variability at the 0.7$_{-0.5}^{+0.7}$\,\% level over a time-scale of 1013 days. 

There are a handful of objects in the literature with previous detections of a change in their dusty emission, as discussed in Section \ref{sources}. For WD J0959$-$0200, \citet{xu2014drop} found a drop in \textit{K} band flux from $57\,\mu$Jy to $46\,\mu$Jy which is equivalent to a peak-to-peak drop of 18.5\%, and \citet{xu2018infrared} found a \textit{K} band peak-to-peak drop of 13\% for WD 1226+110. We note the difficulty in comparing the peak-to-peak variability quoted for these objects with the median variability quoted in this work. With two measurements it is not possible to constrain the distribution of fluxes that an object could be demonstrating, so the quoted variability is peak-to-peak. As we have more \textit{K} band measurements, we use a median variability for all white dwarfs, and compare this to the distribution of the field objects to constrain the level of variability. Typically, peak-to-peak variability is larger than the medium variability that we adopted here. The peak-to-peak variability was also calculated for our sample using the maximum difference between the mean flux at each observation date, and the conclusions are consistent. In this sample we are able to rule out equivalent peak-to-peak variability and median variability for all white dwarfs during the time-scales observed. Our observations of both WD J0959$-$0200 and WD 1226+110 suggest that we are capable of detecting the same level of variability as observed previously, and yet we detect no such variability during our survey time span for all white dwarfs.

\section{Discussion} \label{Discussion}

\subsection{K Band Excess} \label{Phot_Fits}

For all objects in the sample, a white dwarf atmospheric model was fitted to find the expected contribution from the white dwarf photosphere to the \textit{K} band flux. The white dwarf model atmospheres were kindly provided by P. Dufour using models from the Montreal white dwarf group \citep{bergeron2011comprehensive,coutu2019analysis}. The temperature, and log(g) used for the models are shown in Table \ref{tab:WDs}; where possible, these parameters were taken from a previous photometric fit. For a number of white dwarfs where no photometric fit was available, we performed a new fit including data from the \textit{Gaia} Data Release 2  \citep{prusti2016gaia,brown2018gaia}. The white dwarf radii ($R$) were determined from the photometric fit and the distances ($D$) were from the \textit{Gaia} parallax (except for WD J0959$-$0200 which was from \citet{farihi2012trio}). The SED fits are shown in Fig. B1 (available online); the photometric fit works well in most cases. Sometimes, additional adjustments were needed in $R/D$ to improve the fits to the Sloan Digital Sky Survey (SDSS) and Pan-STARRS photometry. Those systems often had a large uncertainty in the parallax.

The white dwarf models were convolved with the bandpass of the UKIRT WFCAM \textit{K} band filter \citep{hewett2006ukirt} to find the expected flux contribution from the white dwarf photosphere. From this, the excess flux associated with the dust was determined. 

White dwarfs are usually photometrically stable and are often used as flux standards. A few white dwarfs in our sample are variable pulsators, however the optical variability is small, typically $<$\,1\%. Assuming the white dwarf photosphere displays no flux variations, we can convert the \textit{K} band variability limit to the dust variability limit, 

\begin{equation}
\sigma _{\textrm{dust}} = \sigma _{\textrm{K}} \, \frac{F_{\textrm{K}}}{F_{\textrm{K}} - F_{\textrm{WD}}},
\label{eq:1}
\end{equation}
where $\sigma _{\textrm{dust}}$ is the variability constraint for the dust, $\sigma _{\textrm{K}}$ is the variability constraint for the \textit{K} band (dust and photosphere), $F_{\textrm{K}}$ is the UKIRT \textit{K} band flux measurement, and $F_{\textrm{WD}}$ is the convolved model flux in the \textit{K} band for the white dwarf photosphere. This puts a constraint on how much the dust could have changed without being detected in this survey assuming the star remains constant.

\begin{figure*}
\centering
\subfloat[\textit{K} band variability]{
  \includegraphics[width=0.48\textwidth]{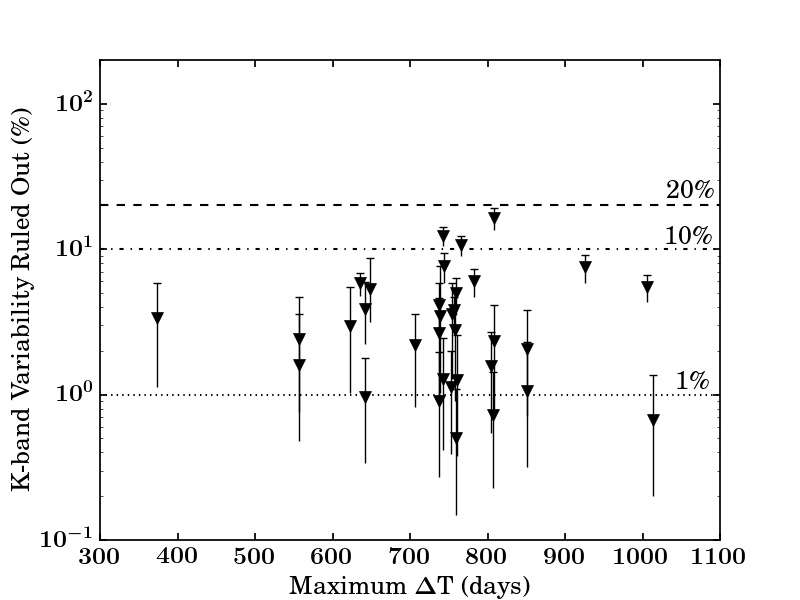}
  \label{fig:Timescale_Plot_2}
}
\hspace{0.1cm}
\subfloat[Dust variability]{
  \includegraphics[width=0.48\textwidth]{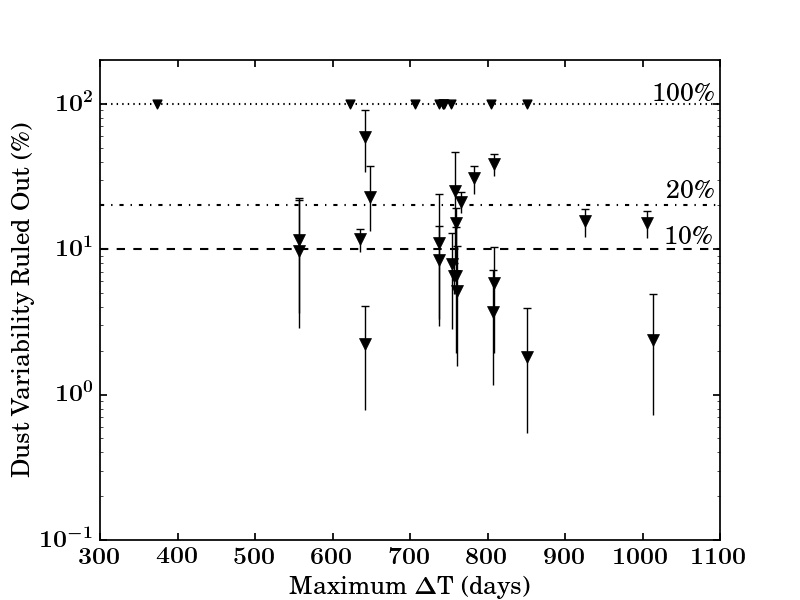}
   \label{fig:Disc_Var}
}
\caption{(a) The maximum \textit{K} band median variability of each white dwarf that would not have been detected over the time-scale of the survey. (b) The maximum dust variability of each white dwarf that would not have been detected over the time-scale of the survey. The variability constraint of the dust component to the \textit{K} band flux was calculated using equation \ref{eq:1}.}

\end{figure*}

We use the maximum variability not detected in the \textit{K} band and convert it to a maximum variability not detected in the dusty emission using equation \ref{eq:1}. Fig. \ref{fig:Disc_Var} is a similar plot to Fig. \ref{fig:Timescale_Plot_2}, but instead showing the level of variability we ruled out for the dusty emission. A \textit{K} band excess is seen for 23/34 of the white dwarfs, as seen in the SEDs in Fig. B1 (available online). The 11 white dwarfs without a significant \textit{K} band excess are: WD 0106$-$328, WD 0300$-$013, WD 0307+077, WD 1041+092, WD 1225$-$079, WD 1456+298, WD 2132+096, WD 2221$-$165, WD 2328+107, WD 1504+329, and WD 1551+175. For the 23 objects with significant \textit{K} band excess, we put limits on how much dust variation could be hidden from detection. For 11 white dwarfs ($\sim$\,50\% of the sample with significant \textit{K} band excesses)  we constrained dust changes to be less than 10\%, demonstrating the dust emission was stable over the course of the monitoring period. For the 11 systems where the \textit{K} band excess is negligible, the dust variability constraint would have been 100\%. For these objects we were only sensitive to an increase in dust emission, not a decrease. Therefore, the dust could have completely disappeared, and we could not distinguish this from the dust displaying no variability. For those with no significant \textit{K} band excess, it is imperative to go to the mid-infrared for dust variability searches.

\subsection{Expected Variability at 4.5\,$\micron$} \label{Spitzer}

Using the variability constraint derived from our UKIRT survey, we can predict the expected variability at 4.5\,$\micron$ by assuming the dust emission has a constant colour temperature (i.e., the change is due to the amount of dust, not the location). So the shape of the dusty emission is assumed to remain constant, such that the percentage change of the dusty emission at \textit{K} wavlengths can be equalled to the percentage change at 4.5\,$\micron$. Paralleling the \textit{K} band analysis and equation 1, we can calculate the expected variability at \textit{Spitzer} 4.5\,$\micron$ wavelengths,
\begin{equation}
\sigma _{\textrm{[4.5]}} = \sigma _{\textrm{dust}} \, \frac{F_{\textrm{[4.5]}}  - F_{\textrm{WD}}}{F_{\textrm{[4.5] }} },
\label{eq:2}
\end{equation}
where $\sigma _{\textrm{[4.5]}}$ is the predicted variability constraint for the 4.5\,$\micron$ \textit{Spitzer} waveband, $\sigma _{\textrm{dust}}$ is calculated from the \textit{K} band using equation \ref{eq:1}, $F_{\textrm{[4.5]}}$ is the 4.5\,$\micron$ \textit{Spitzer} flux, and $ F_{\textrm{WD}}$ is the convolved model flux in the 4.5\,$\micron$ band for the white dwarf photosphere.

\begin{figure}
   \centering 
   \captionsetup{font=footnotesize}
   \includegraphics[width=0.48\textwidth]{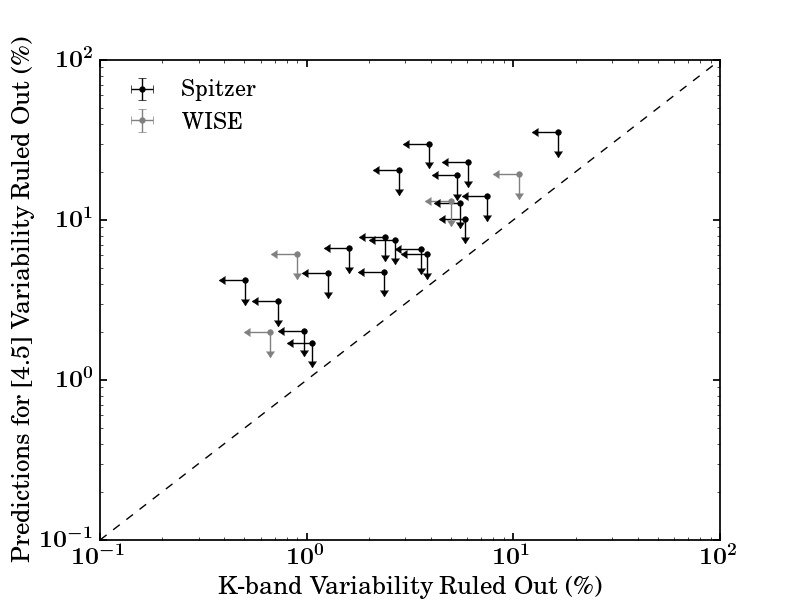}
   \caption{Predictions for the maximum median variability at 4.5\,$\micron$ that would have escaped detection based on the maximum median \textit{K} band variability not detected with this UKIRT survey. The arrows represent each axis is a constraint on the level of variability. Most objects had \textit{Spitzer} 4.5\,$\micron$ data, but for four objects \textit{WISE} data was used as no \textit{Spitzer} data was available. }   
  
  \label{fig:Spitzer_Comparison}
\end{figure}

Fig. \ref{fig:Spitzer_Comparison} shows the \textit{K} band variability constraint, converted into a \textit{Spitzer} 4.5\,$\micron$ band variability constraint using equation \ref{eq:2}. The \textit{Spitzer} data was taken from the references listed in Table \ref{tab:WDs}. If there was no \textit{Spitzer} 4.5\,$\micron$ data available then \textit{WISE} 4.6\,$\micron$ data was used instead. The dust excess at 4.5\,$\micron$ is larger than that at the \textit{K} band, therefore, the variability constraint in flux units will be larger for 4.5\,$\micron$, as seen in the figure. Those white dwarfs with negligible \textit{K} band excess are excluded from this study. For those 23 white dwarfs with significant \textit{K} band excess, at \textit{Spitzer} wavelengths we predict that variability can be ruled out above 10\% for $\sim$60\% of our sample, above 20\% for $\sim$\,85\% of our sample, and above 36\% for all our sample. 
 
\begin{figure}
   \centering 
   \captionsetup{font=footnotesize}
   \includegraphics[width=0.5\textwidth]{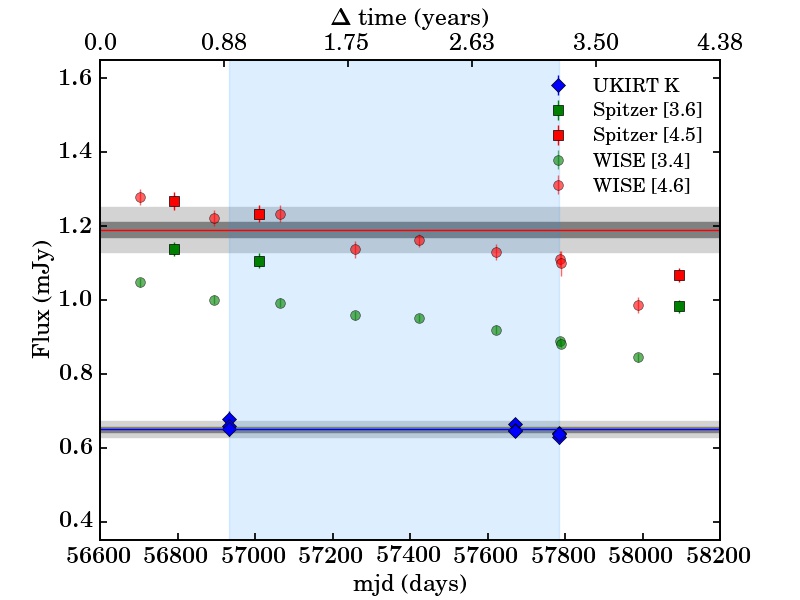}
   \caption{A light curve of WD 0408$-$041 showing the UKIRT \textit{K} band data, and the \textit{Spitzer}/\textit{WISE} data at which \citet{farihi2018dust} observed a drop in the mid-infrared flux. The grey regions represent the 1 and 3\,$\sigma$ median variability constraint for the \textit{K} band, and then scaled up to \textit{Spitzer} wavelengths using equations \ref{eq:1} and \ref{eq:2} (assuming a constant colour temperature). The blue region shows the period of time that we have UKIRT observations. Figure adapted from \citet{farihi2018dust}.}
  \label{fig:GD56_LC}
\end{figure}

A particular application using our survey can be applied to WD 0408$-$041 system. \citet{farihi2018dust} found the dust emission of WD 0408$-$041 to increase and decrease with a peak-to-peak variability of 20\% over $\sim$\,11 years. As can be seen from Fig. B1 (available online), the SED of WD 0408$-$041 demonstrates a large \textit{K} band excess. Therefore, any variability observed at mid-infrared wavelengths should be mimicked at a lower level in the \textit{K} band, unless there is a colour change associated with the flux change. Fig. \ref{fig:GD56_LC} shows the light curve for WD 0408$-$041 over approximately four years showing \textit{Spitzer}, \textit{WISE} and UKIRT \textit{K} band data. The drop in mid-infrared flux is evident. We do not have UKIRT \textit{K} band observations which cover the full mid-infrared time-scale, as shown by the blue shaded region. Using equations \ref{eq:1} and \ref{eq:2}, the \textit{K} band variability we ruled out using the UKIRT survey is scaled up to mid-infrared wavelengths. The change in the \textit{Spitzer} 4.5\,$\micron$ flux is consistent with the variability constraint derived from the \textit{K} band data, implying that the variability could be consistent with the same colour temperature of the dust emission during the three year UKIRT survey. However, if we look across the four year baseline, even at 3\,$\sigma$, the \textit{Spitzer} flux values lie outside of the shaded region, i.e. prior to 56880 MJD and after 58100 MJD. The most significant drop in \textit{Spitzer} flux was around 58100 days. We do not have simultaneous \textit{K} band observations, with the closest \textit{K} band measurement being over a year previous to this. There is previous evidence that large scale flux changes can occur on time-scales of less than a year \citep{xu2014drop}. A further \textit{K} band measurement is needed in order to understand whether the \textit{K} band flux also dropped around MJD58100, or if there is a colour change associated with this variability.

\subsection{Variability Time-scales} \label{timescales-section}

\begin{figure}
   \centering 
   \captionsetup{font=footnotesize}
   \includegraphics[width=0.48\textwidth]{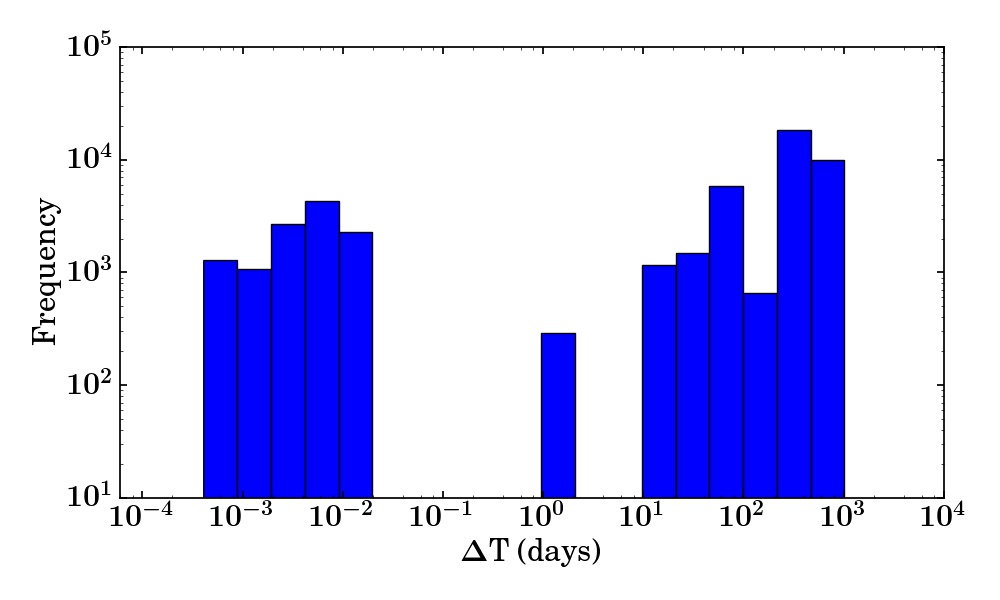}
   \caption{A histogram showing the characteristic survey sampling to search for variability. The plot shows the distribution of all time-scales between frames of a given white dwarf, for all white dwarfs. This demonstrates that for all white dwarfs our survey was sensitive to minute/hour time-scales and year time-scales.} 
  
  \label{fig:Timescale_Plot}
\end{figure}

Fig. \ref{fig:Timescale_Plot} shows the characteristic sampling for all the white dwarfs in the sample. This survey was sensitive to short time-scales on the order of minutes to hours and long time-scales on the order of years. Previous examples of gas variability show changes on both these time-scales, and dust variability show changes on year long time-scales \citep[e.g.][]{xu2014drop, manser2019planetesimal}.

\begin{figure}
   \centering 
   \captionsetup{font=footnotesize}
   \includegraphics[width=0.5\textwidth]{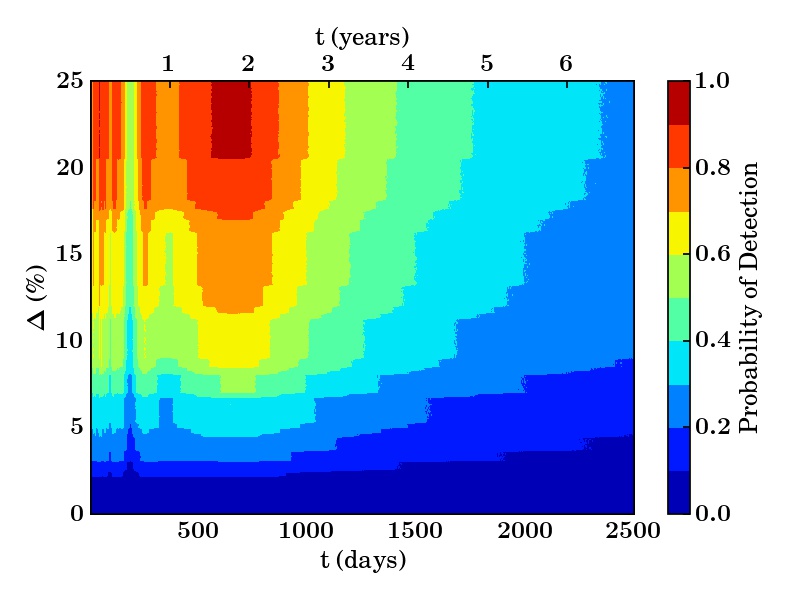}
   \caption{A toy model in which every white dwarf in our sample is considered to have a \textit{K} band flux that varies by $\Delta \%$ every $t$ days. The density plot indicates the probability of detecting that combination of $\Delta$ and $t$ for white dwarfs in the UKIRT survey. The smallest grid scale is 1 day and 0.1\%.}
  \label{fig:model_plot}
\end{figure}

We considered a toy model where the \textit{K} band flux of every white dwarf varies as a step function by $\Delta$\%, every $t$ days. We calculated the probability of detecting this variability in this UKIRT survey. Each white dwarf has a minimum detectable variability which is 3\,$\sigma$ above the median variability of the field stars at the same magnitude as the white dwarf, where $\sigma$ is the standard deviation on this median variability. If $\Delta$ was smaller than the minimum detectable variability for that white dwarf, the probability of detection was 0. For every combination of $\Delta$ and $t$, the light curve was randomly sampled 1000 times for each white dwarf using the UKIRT sampling times to calculate the probability that if it were varying in this manner, it would have been detected in this survey. By averaging over all 34 white dwarfs at each $\Delta$ and $t$, the overall probability of detection was deduced, as shown in Fig. \ref{fig:model_plot}. The colour scheme represents the probability that the UKIRT survey would have detected that combination of $\Delta$ and $t$ if all white dwarfs varied in this way. The highest probability time-scales show that our survey was sensitive to variability over days, and years. We can rule out models where all white dwarfs vary on time-scales of $\sim$\,1--2 years with large amplitudes. However, we cannot rule out models where white dwarfs vary on time-scales much longer than this, or with smaller amplitudes ($<$\,10\%). This may explain why some objects from the literature which have been previously found to vary, are consistent with little or no variability in this survey. For example, using seven years of \textit{WISE} data, \citet{swan2019most} found variability in a number of dusty white dwarfs. Using this model we can see that for these longer time-scales, we have less than a 30\% chance of detecting variability.

\subsection{Implications}

The leading explanation for white dwarf pollution cites asteroids as the source which are scattered inwards stochastically. How frequently these asteroids are scattered inwards remains an open question, with debate in the literature between whether pollution is caused by a single large body, or many small bodies \citep[e.g.][]{jura2008pollution, jura2009x, wyatt2014stochastic}. 

\citet{wyatt2014stochastic} presented a model whereby white dwarf pollution is explained by a continuous stream of many small asteroids, with the stochastic scattering of the largest bodies dominating the accretion rates. In this model, the lifetime of the dusty material is a free parameter, with a best-fitting value of 20 years. Our survey would be unlikely to detect variations on such long time-scales. However, if the disc lifetime was instead an order of magnitude smaller ($\sim$\,2 years), large amplitude variations in the infrared flux on these time-scales would be expected. In our survey, we rule out large variability over these time-scales. Therefore, the results of this survey provide further evidence in support of long ($>$\,3 year) lifetimes for the dusty material around white dwarfs. It also demonstrates that disruption events large enough to be detectable in the infrared do not regularly occur on short time-scales such as years. This is crucial, as for many highly polluted DA white dwarfs, material can sink out of the atmospheres on time-scales of days to years. This work provides observational evidence that dusty material has a lifetime of at least 2--3 years. Therefore, we highlight the need for further monitoring in the infrared to discover the frequency of scattering events.

\section{Conclusions} \label{Conclusion}

Infrared studies are crucial for understanding the accretion processes associated with polluted white dwarfs. This work reports a near-infrared \textit{J}, \textit{H}, and \textit{K} band monitoring campaign with the UKIRT, completed over three years which studied 34 polluted white dwarfs with infrared excesses. High precision \textit{J}, \textit{H}, and \textit{K} band photometry was reported for the white dwarfs in the sample, for many of these objects high precision photometry is not previously reported. The main results are summarised as follows.

\begin{enumerate}

  \item The \textit{K} band flux was stable within the error of the survey for all observations of the white dwarfs, where the observation sampling rates were on the order of minutes and years. Of the near-infrared bands, the \textit{K} band is the most sensitive to the dust. We ruled out median \textit{K} band variability greater than 16.4$^{+2.8}_{-2.8}$\% for the faintest white dwarf in the sample and less than 2.88$^{+2.5}_{-1.9}$\% for half of our sample. For $>$\,90\% of our sample we ruled out variability at the 10\% level, and for our 10 most constrained white dwarfs we ruled out \textit{K} band variability at the percent level, $<$\,1.57$^{+1.1}_{-1.0}$\%. 
  
    \item WD J0959$-$0200 and WD 1226+110 have previously exhibited variability in their dusty emission as seen at the \textit{K} band (18.5\% for WD J0959$-$0200 and 13\% for WD 1226+110) and \textit{Spitzer} 3.6 and 4.5\,$\micron$ bands. These objects did not display any significant variability in our study. For most of our sample (except our faintest object, WD 1554+094) we ruled out similar \textit{K} band variability to these white dwarfs. This highlights that these large  \textit{K} band variability events are rare.
     
   \item Assuming the dust emission does not change its colour temperature, we predicted variability at 4.5\,$\micron$ based on the \textit{K} band variability. We found that the white dwarfs should not vary at 4.5\,$\micron$ above 10\% for $\sim$\,60\% of our sample, above 20\% for $\sim$\,85\% of our sample, and above 36\% for all our sample. 
   
  \item We demonstrated that any white dwarf with large amplitude ($>$\,20\%) near-infrared variability on short ($<$\,3 year) time-scales should have been found with this survey. This implies that major tidal disruption events that lead to white dwarf pollution occur less frequently than every three years, and are consequently rare. Infrared variability around white dwarfs with infrared excesses does exist as revealed by other observations. The constraints provided from our \textit{K} band survey imply that such events, if they occur, do so on short time-scales, and quickly become stable within a few years. 
  
  \item Ground based near-infrared searches for white dwarf dust variability are fruitful, and as demonstrated in this survey can search for variability down to the percent level for the brightest sources. This survey was not sensitive to long time-scale ($>$\,3 year) variability and, therefore, we highlight the need for long-term monitoring of white dwarfs with infrared excesses to fully understand these highly enigmatic and dynamic systems. 

\end{enumerate}

%%%%%%%%%%%%%%%%%%%%%%%%%%%%%%%%%%%%%%%%%%%%%%%%%%

\section*{Acknowledgements}

We would like to thank Matthew Auger, Patrick Dufour, Bruce Gary, John Harrison, Jonathan Irwin, Mike Irwin and Mark Wyatt for useful contributions and discussions which helped shape the paper. We would like to thank the anonymous referee for their comments which improved the manuscript. LR would like to acknowledge funding from The Science and Technology Facilities Council, Jesus College (Cambridge), and The University of Cambridge. AB would like to acknowledge funding from the Royal Society Dorothy Hodgkin Fellowship. TvH was supported by the National Science Foundation Award AST-1715718. This work is partly supported by the Gemini Observatory, which is operated by the Association of Universities for Research in Astronomy, Inc., on behalf of the international Gemini partnership of Argentina, Brazil, Canada, Chile, the Republic of Korea, and the United States of America.

We gratefully acknowledge data obtained from the UKIRT. The United Kingdom Infrared Telescope is operated by the Joint Astronomy Centre on behalf of the U.K. Particle Physics and Astronomy Research Council. This work has made use of the Montreal White Dwarf Database (http://dev.montrealwhitedwarfdatabase.org) \citep{dufour2017mwdd}. This work has made use of data from the European Space Agency (ESA) mission
{\it Gaia} (\url{https://www.cosmos.esa.int/gaia}), processed by the {\it Gaia}
Data Processing and Analysis Consortium (DPAC,
\url{https://www.cosmos.esa.int/web/gaia/dpac/consortium}). Funding for the DPAC has been provided by national institutions, in particular the institutions
participating in the {\it Gaia} Multilateral Agreement.

%%%%%%%%%%%%%%%%%%%%%%%%%%%%%%%%%%%%%%%%%%%%%%%%%%



\bibliographystyle{mnras}
\bibliography{refs_3} % if your bibtex file is called example.bib

%%%%%%%%%%%%%%%%%%%%%%%%%%%%%%%%%%%%%%%%%%%%%%%%%%

%%%%%%%%%%%%%%%%% APPENDICES %%%%%%%%%%%%%%%%%%%%%

\appendix

\section{Observation Information}

Table \ref{tab:DATES} shows the dates of the observations for each white dwarf in the sample. Table \ref{tab:Frames} shows the frame numbers for each white dwarf in the sample.

\begin{table*}
	\centering
	\caption{All the polluted white dwarfs with an infrared excess observed with the UKIRT and the dates they were observed.}
	\label{tab:DATES}
	\begin{tabular}{lccccccccc} 
		\hline
		WD Name & 2014B & 2015A & 2015B & 2016A & 2016A & 2016B & 2016B & 2017A & 2017A \\
		\hline

WD 0010+280 & 03/10/14 & N & N & N & N & 06/07/16 & N & N & N \\
WD 0106$-$328 & 03/10/14 & N & N & N & N & 11/10/16 & N & N & N \\
WD 0146+187 & 03/10/14 & N & N & N & N & 12/07/16 & N & N & N \\
WD 0300$-$013 & 03/10/14 & N & N & N & N & 22/07/16 & 31/01/17 & N & N \\
WD 0307+077 & 03/10/14 & N & N & N & N & 10/10/16 & N & N \\
WD 0408$-$041 & 03/10/14 & N & N & N & N & 10/10/16 & 31/01/17 & N & N \\
WD 0435+410 & 03/10/14 & N & 31/01/16 & N & N & 10/10/16 & N & N & N \\
WD J0738+1835 & N & 03/05/15 & N & 29/02/16 & N & 10/10/16 & 28/01/17 & N & N \\
WD J0959$-$0200 & 09/11/14 & 02/02/15 & N & 05/03/16 & 07/05/16 & N & N & 23/05/17 & N \\
WD 1015+161 & 30/10/14 & N & N & 09/05/16 & N & N & N & N & N \\
WD 1018+410 & 30/10/14 & N & N & 09/05/16 & N & N & N & N & N \\
WD 1041+092 & N & 02/05/15 & N & 04/03/16 & 09/05/16 & N & N & N & N \\
WD 1116+026 & N & 02/05/15 & N & 05/03/16 & 10/05/16 & N & N & 30/05/17 & N \\
WD 1145+017 & N & 02/05/15 & N & N & N & N & N & 09/05/17 & 24/05/17* \\
WD 1145+288 & N & 02/05/15 & N & 23/02/16 & 09/05/16 & N & N & 30/05/17 & N \\
WD 1150$-$153 & N & 03/05/15 & N & 05/03/16 & 10/05/16 & N & N & 01/06/17 & N \\
WD J1221+1245 & N & 02/05/15 & N & 01/03/16 & 10/05/16 & N & N & 25/05/17 & N \\
WD 1225$-$079 & N & 03/05/15 & N & 04/03/16 & 10/05/16 & N & N & 25/05/17 & N \\
WD 1226+110 & N & 02/05/15 & N & 29/02/16 & 10/05/16 & N & N & 09/05/17 & N \\
WD J1234+5606 & N & 03/05/15 & N & 04/03/16 & 05/06/16 & N & N & 07/06/17 & N \\
WD 1349$-$230 & N & 03/05/15 & N & 04/03/16 & 06/06/16 & N & N & 30/05/17 & N \\
WD 1456+298 & N & 02/05/15 & N & 17/02/16 & 10/05/16 & N & N & 14/05/17 & N \\
WD 1504+329 & N & 01/05/15 & N & 29/02/16 & 08/05/16 & N & N & 12/05/17 & N \\
WD 1536+520 & N & 02/05/15 & N & 05/03/16 & 29/05/16 & N & N & 28/05/17 & N \\
WD 1551+175 & N & 01/05/15 & N & 03/03/16 & 31/05/16 & N & N & 14/05/17 & N \\
WD 1554+094 & N & 01/05/15 & N & 29/02/16 & 08/05/16 & N & N & 20/05/17 & 18/07/17 \\
WD J1617+1620 & N & 01/05/15 & N & 17/02/16 & 10/05/16 & N & N & 10/05/17 & 17/07/17 \\
WD 1729+371 & N & 02/05/15 & 02/08/15 & 28/02/16 & 10/05/16 & N & N & 12/05/17 & 17/07/17 \\
WD 1929+011 & 06/10/14 & N & 02/08/15 & N & N & 10/07/16 & 10/05/17 & 15/07/17 & N \\
WD 2132+096 & N & 03/05/15 & 09/08/15 & 21/04/16 & N & 10/07/16 & N & 11/05/17 & 15/07/17 \\
WD 2207+121 & 13/10/14 & N & 08/08/15 & N & N & 29/06/16 & N & 15/07/17 & N \\
WD 2221$-$165 & 30/10/14 & N & 09/08/15 & N & N & 13/07/16 & N & N & N \\
WD 2326+049 & 03/10/14 & N & 09/08/15 & N & N & 06/07/16 & N & N & N \\
WD 2328+107 & N & N & 08/08/15 & 01/06/16 & N & 06/07/16 & N & 02/06/17 & 15/07/17 \\

		\hline
	\end{tabular}
	\begin{tablenotes}
	 \item (*) For WD 1145+017, there are three more dates for 2017A: 06/06/2017, 21/06/2017 and 22/06/2017. 
	\end{tablenotes}	
	
\end{table*}

\begin{table*}
	\centering
	\caption{All the polluted white dwarfs with an infrared excess observed with the UKIRT survey and the total number of frames. Each frame consists of a dithered stack of 5 exposures, each with the corresponding frame time of 5 or 10 seconds. FT = frame time for each individual exposure in the dithered stack, FN = total number of dithered stacks over all observations.}
	\label{tab:Frames}
	\begin{tabular}{lcccccc} 
		\hline
		WD Name & FT J (secs) & FN J & FT H (secs) & FN H & FT K (secs) & FN K \\
		\hline

WD 0010+280 & 5 & 6 & 10 & 6 & 10 & 10 \\
WD 0106$-$328 & 5 & 6 & 5 & 6 & 5 & 6 \\
WD 0146+187 & 5 & 6 & 5 & 6 & 5 & 6 \\
WD 0300$-$013 & 5 & 6 & 5 & 6 & 5 & 9 \\
WD 0307+077 & 5 & 6 & 10 & 6 & 10 & 10 \\
WD 0408$-$041 & 5 & 9 & 5 & 9 & 5 & 9 \\
WD 0435+410 & 5 & 3 & 5 & 3 & 5 & 9 \\
WD J0738+1835 & 10 & 9 & 10 & 27 & 10 & 75 \\
WD J0959$-$0200 & 10 & 9 & 10 & 27 & 10 & 75 \\
WD 1015+161 & 5 & 6 & 5 & 6 & 10 & 6 \\
WD 1018+410 & 5 & 6 & 10 & 6 & 10 & 10 \\
WD 1041+092 & 10 & 9 & 10 & 21 & 10 & 51 \\
WD 1116+026 & 5 & 4 & 5 & 4 & 5 & 12 \\
WD 1145+017 & 10 & 18 & 10 & 42 & 10 & 102 \\
WD 1145+288 & 10 & 12 & 10 & 28 & 10 & 68 \\
WD 1150$-$153 & 5 & 12 & 5 & 11 & 10 & 12 \\
WD J1221+1245 & 10 & 12 & 10 & 36 & 10 & 100 \\
WD 1225$-$079 & 5 & 4 & 5 & 4 & 5 & 12 \\
WD 1226+110 & 5 & 12 & 10 & 12 & 10 & 20 \\
WD J1234+5606 & 10 & 9 & 10 & 27 & 10 & 75 \\
WD 1349$-$230 & 5 & 12 & 10 & 12 & 10 & 20 \\
WD 1456+298 & 5 & 3 & 5 & 3 & 5 & 6 \\
WD 1504+329 & 10 & 9 & 10 & 27 & 10 & 75 \\
WD 1536+520 & 10 & 12 & 10 & 28 & 10 & 68 \\
WD 1551+175 & 10 & 12 & 10 & 28 & 10 & 68 \\
WD 1554+094 & 10 & 15 & 10 & 45 & 10 & 125 \\
WD J1617+1620 & 5 & 19 & 10 & 23 & 10 & 45 \\
WD 1729+371 & 5 & 19 & 5 & 19 & 10 & 23 \\
WD 1929+011 & 5 & 6 & 5 & 10 & 5 & 20 \\
WD 2132+096 & 5 & 20 & 5 & 22 & 10 & 26 \\
WD 2207+121 & 10 & 18 & 10 & 36 & 10 & 76 \\
WD 2221$-$165 & 5 & 6 & 5 & 6 & 5 & 6 \\
WD 2326+049 & 5 & 3 & 5 & 3 & 5 & 9 \\
WD 2328+107 & 5 & 19 & 5 & 19 & 5 & 23 \\

		\hline
	\end{tabular}
\end{table*}

\end{CJK}
% Don't change these lines
\bsp	% typesetting comment
\label{lastpage}
\clearpage
\includepdf[pages = 1-5]{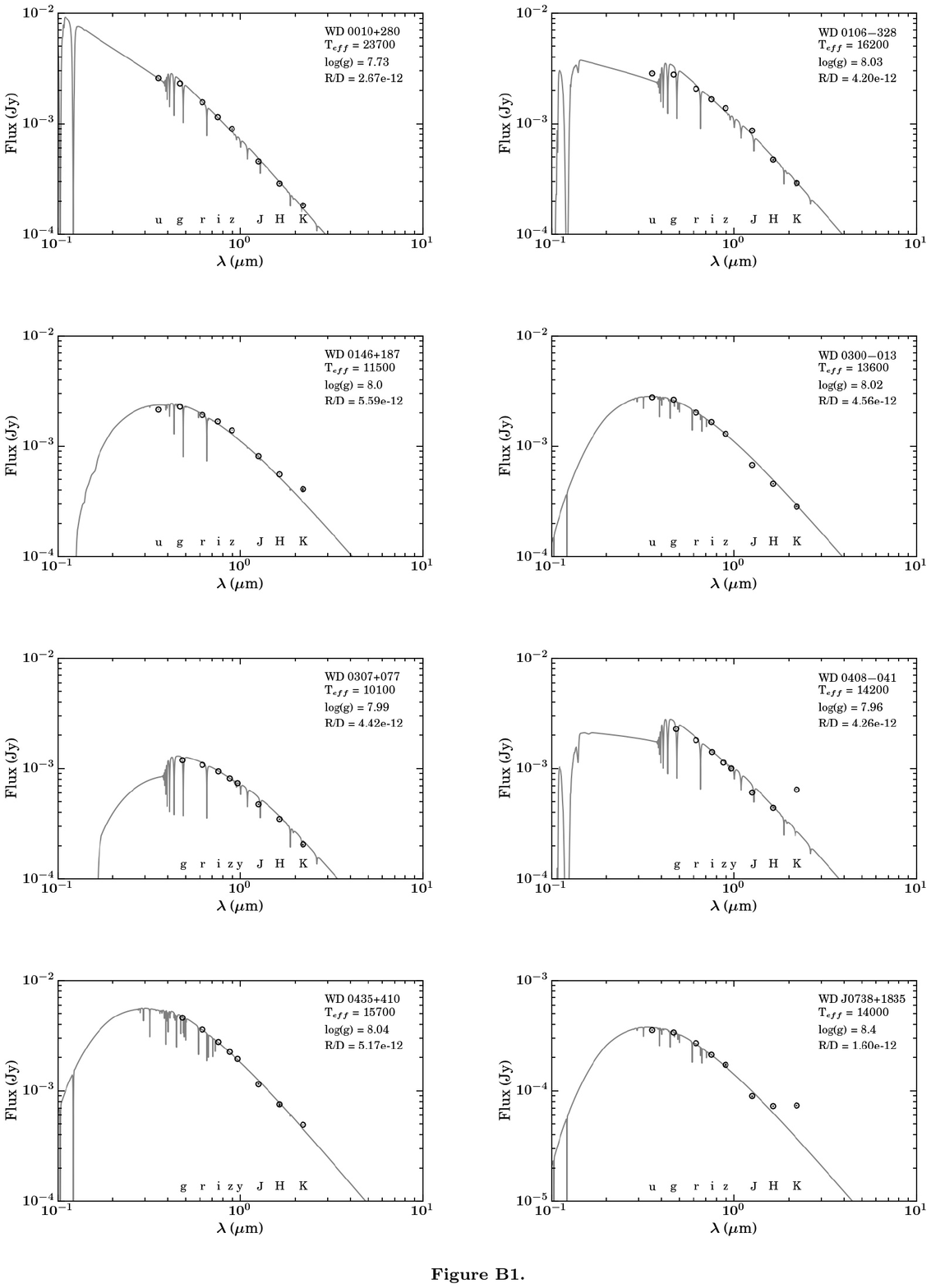} 

\end{document}